\documentclass[letterpaper, 10 pt, conference,dvipsnames]{ieeeconf}  

\IEEEoverridecommandlockouts                              

\usepackage{amsmath,amssymb,mathtools}
\usepackage{xcolor}
\usepackage{graphicx}
\usepackage{tikz}
\usepackage{pgfplots}
\usepackage{pgf}
\usepgfplotslibrary{colorbrewer}
\usepackage[font=small,skip=0.1pt]{caption}
\usepackage{todonotes}

\DeclareMathOperator{\A}{\mathcal{A}}
\DeclareMathOperator{\B}{\mathcal{B}}
\DeclareMathOperator{\R}{\mathbb{R}}
\DeclareMathOperator{\Han}{\mathfrak{H}}

\newtheorem{lemma}{Lemma}
\newtheorem{definition}{Definition}
\newtheorem{theorem}{Theorem}
\newtheorem{corollary}{Corollary}
\newtheorem{remark}{Remark}

\title{\LARGE \bf
From System Level Synthesis to Robust Closed-loop Data-Enabled Predictive Control
}

\author{Yingzhao Lian and Colin N. Jones
\thanks{This work has received support from the Swiss National Science Foundation under the RISK project (Risk Aware Data-Driven Demand Response), grant number 200021 175627.}
\thanks{Yingzhao Lian and Colin N. Jones are with Automatic Laboratory, Ecole Polytechnique Federale de Lausanne, Switzerland.
        {\tt\small
        $\{$yingzhao.lian, colin.jones$\}$@epfl.ch}}%
}

\begin{document}

\maketitle
\thispagestyle{empty}
\pagestyle{empty}

\begin{abstract}
Willems' fundamental lemma and system level synthesis both characterize a linear dynamic system by its input/output sequences. In this work, we extend the application of the fundamental lemma from deterministic to uncertain LTI systems and then further prove this extension to be equivalent to system level synthesis. Based on this uncertain extension, a robust closed-loop data-enabled predictive control scheme is proposed, where a causal feedback control law is further derived. Two numerical experiments, including the temperature control of a single-zone building, are carried out to validate the effectiveness of the proposed data-driven controller.

\end{abstract}

\section{Introduction}
Although linear control theory is well-developed, the Willems' fundamental lemma~\cite{willems2005note} and the system level synthesis~\cite{anderson2019system} still spark significant research interest. The Willems' fundamental lemma enables a data-driven characterization of a deterministic linear time invariant (LTI) system under reasonable assumptions on controllability and persistent excitation. This result has been successfully applied in controller design~\cite{coulson2019data,de2019formulas,markovsky2007linear,markovsky2008data}, which shows better data efficiency than, for example, reinforcement learning. Further results in~\cite{yu2021controllability} relax the condition of the fundamental lemma, which improves the scalability of Willems' fundamental lemma in practical applications. Beyond the scope of LTI systems, \cite{lian2021nonlinear,bisoffi2020data,berberich2020trajectory,rueda2020data} attempt to extend the lemma to various nonlinear systems.

Different from Willems' fundamental lemma, system level synthesis (SLS)~\cite{anderson2019system} is a model-based framework built for general uncertain linear systems, which are not necessarily time-invariant. The main benefit of the SLS scheme is that it provides an explicit link between a system's response and a linear feedback control law. As a model-based method, a direct integration of SLS into well-developed linear control theory allows the consideration of model structure and parametric uncertainty~\cite{lamperski2013output,tanaka2014optimal}.

The link between these two methods motivates this work. In~\cite[Theorem 2]{xue2020data}, a link between SLS and the fundamental lemma is derived for deterministic LTI systems. In this work, we further generalize this result to uncertain LTI systems in Lemma~\ref{lem:eqv}. Inspired by this observation, we develop a causal robust closed-loop data-enabled predictive control scheme, which results in the same scale of computational cost as a model-based robust MPC with a linear feedback law and full state measurements. The major contributions of this work are summarized as follows:
\begin{itemize}
    \item Show the equivalence between the Willems' fundamental lemma and the SLS for uncertain LTI systems.
    \item A robust closed-loop data-enabled predictive control scheme is proposed with a causal feedback structure.
\end{itemize}

\subsection*{\textbf{Notation}}
$\text{colspan}(A)$ denotes the column space (\textit{e.g.} range) of the matrix A. $I$ is the identity matrix and $\textbf{O}$ is a matrix of all zeros. $x: = \{x_i\}_{i=0}^T$ denotes a set of size $T$ indexed by $i$. $x_i$ denotes the measurement of $x$ at time $i$, a boldface $\textbf{x}:=[x_0^\top,x_1^\top\dots x_L^\top]^\top$ denotes a concatenated sequence of $x_i$. Meanwhile, the $\Tilde{\enspace}$ sign indicates that a measured datapoint lies within the dataset. For the sake of consistency, $L$ is reserved for the length of the system responses and $n_c$ denotes the number of columns of a Hankel matrix.

\section{Preliminaries}\label{sect:pre}
\subsection{Willems' Fundamental Lemma}
\begin{definition}
A Hankel matrix of depth $L$ associated with a signal sequence $\{s_i\}_{i=0}^T,\;s_i\in\R^{n_s}$ is
\begin{align*}
    \Han_L(s):=
    \begin{bmatrix}
    s_0 & s_1&\dots&s_{T-L}\\
    s_1 & s_2&\dots&s_{T-L+1}\\
    \vdots &\vdots&&\vdots\\
    s_{L-1} & s_{L}&\dots&s_T
    \end{bmatrix}\;.
\end{align*}
\end{definition}

A deterministic LTI system, dubbed $\mathfrak{B}(A,B,C,D)$, is defined as
\begin{align}\label{eqn:lin_dyn_deter}
\begin{split}
    x_{i+1} = Ax_i+Bu_i\;,\;y_i = Cx_i+Du_i\;,
\end{split}
\end{align}
whose order is denoted by $\mathfrak{D}(\mathfrak{B}(A,B,C,D)):=n_x$. An $L$-step trajectory generated by this system is 
\begin{align*}
    [\textbf{u}^\top,\textbf{y}^\top]:=[u_0^\top,\dots,u_{L-1}^\top,y_{0}^\top,\dots,y_{L-1}^\top]^\top\;.
\end{align*} 
The set of all possible $L$-step trajectories generated by $\mathfrak{B}(A,B,C,D)$ is denoted by $\mathfrak{B}_L(A,B,C,D)$. 

Given a sequence of input-output measurements $\{\Tilde{u}_i,\Tilde{y}_i\}_{i=0}^T$, we call the input sequence persistently exciting of order $L$ if $\Han_L(\Tilde{u})$ is full row rank. By building the following $n_c$-column stacked Hankel matrix 
\begin{align}\label{eqn:fund_column}
    \Han_L(\Tilde{y},\Tilde{u}):=\begin{bmatrix}
    \Han_{L}(\Tilde{u})\\\Han_L(\Tilde{y})
    \end{bmatrix}\;,
\end{align}
we state the \textbf{Fundamental Lemma} as
\begin{lemma}\label{lem:funda}\cite[Theorem 1]{willems2005note}
Consider a controllable linear system and assume $\{\Tilde{u}\}_{i=0}^T$ is persistently exciting of order $L+ \mathfrak{O}(\mathfrak{B}(A,B,C,D)$. Then $\text{colspan}(\Han_L(\Tilde{u},\Tilde{y}))=\mathfrak{B}_L(A,B,C,D)$.
\end{lemma}

\subsection{Data-enabled Predictive Control}
Data-enabled predictive control (DeePC)~\cite{coulson2019data} is a predictive control scheme, which applies the fundamental lemma to enable data-driven prediction for deterministic LTI systems. With a noise-free dataset measured offline, $\Tilde{u}$ and $\Tilde{y}$, a predictive control problem of horizon $n_h$ is formulated as
\begin{align}\label{eqn:deepc}
    \begin{split}
        \min\limits_{\textbf{y}_{pred},\textbf{u}_{pred},g} \;&\; J(\textbf{y}_{pred},\textbf{u}_{pred})\\
        \text{s.t.} \;
        &\;\begin{bmatrix}
    \Han_{L,init}(\Tilde{u})\\\Han_{L,init}(\Tilde{y})\\\Han_{L,pred}(\Tilde{u})\\\Han_{L,pred}(\Tilde{y})
    \end{bmatrix}g=\begin{bmatrix}
        \textbf{u}_{init}\\\textbf{y}_{init}\\\textbf{u}_{pred}\\\textbf{y}_{pred}\\
    \end{bmatrix}\\
        &\; F_u\textbf{u}_{pred}\leq f_u,\; F_y\textbf{y}_{pred}\leq f_y\\
    \end{split}\;,
\end{align}
where $J(\cdot,\cdot)$ is a convex objective function. $F_u, f_u, F_y$ and $f_y$ models polytopic input and output constraints. $\textbf{u}_{init},\textbf{y}_{init}$ are a fixed-length sequence of measured inputs and outputs preceding the current point in time. The matrix $\Han_L(\Tilde{u})$ is split into two sub-Hankel matrices as
\begin{align*}
    \Han_L(\Tilde{u}) = \begin{bmatrix}
        \Han_{L,init}(\Tilde{u})\\\Han_{L,pred}(\Tilde{u})
    \end{bmatrix}\;.
\end{align*}
The matrix $\Han_{L,init}(\Tilde{u})$ is of depth $t_{init}$ and $\Han_{L,pred}(\Tilde{u})$ is of depth $n_h$ such that $t_{init}+n_h = L$. The matrices $\Han_{L,init}(\Tilde{y})$, $\Han_{L,pred}(\Tilde{y})$ are defined accordingly. The choice of $t_{init}$ is made to ensure a unique estimation of the initial state; please refer to~\cite{markovsky2008data} for more details.

\subsection{System Level Synthesis}
For the sake of simplicity, system level synthesis (SLS) is introduced for a fully observed uncertain LTI system and a more general setup can be found in~\cite{anderson2019system}.
\begin{align}\label{eqn:lin_dyn_stc}
    \begin{split}
        x_{i+1} = Ax_{i}+Bu_{i}+w_i\;,
    \end{split}
\end{align}
where $w_i$ is the process noise. To proceed, we further define
\begin{align*}
    \A &:= \text{blkdiag}(\overbrace{A,A,\dots,A}^{L-1},\textbf{O})\;,\\
    \B &:= \text{blkdiag}(\underbrace{B,B,\dots,B}_{L-1},\textbf{O})\;.
\end{align*}

The response of system~\eqref{eqn:lin_dyn_stc} is 
\begin{align*}
    \textbf{x} = Z\A\textbf{x}+Z\B\textbf{u}+\textbf{w}\;,
\end{align*}
where $\textbf{x},\textbf{u},\textbf{w}$ are the sequences of states, inputs and disturbances with $\textbf{x}:=[x_0^\top,x_1^\top\dots x_L^\top]^\top$ and $\textbf{u}$ and $\textbf{w}$ are defined accordingly, where $Z$ is the shift operator. A linear time-varying feedback control law is defined as
\begin{align}\label{eqn:fb}
    \textbf{u} &= \begin{bmatrix}
    K_{0,0}&&&\\
    K_{1,0}&K_{1,1}&&\\
    \vdots&\ddots&\ddots&\\
    K_{T,0}&\dots&K_{L,L-1}&K_{L,L}
    \end{bmatrix}\textbf{x}\ := \textbf{K}\textbf{x}\;,
\end{align}
where $K_{i,j}$ is the feedback law injecting into $u_i$ with respect to the measurement $x_j$.  System level synthesis (SLS) characterizes all trajectories driven by some linear feedback control law and it is stated as 
\begin{theorem}
(\cite[Theorem 2.1]{anderson2019system}) Over a horizon $t = 0, . . . , L-1$, the system dynamics~\eqref{eqn:lin_dyn_stc} with block-lower-triangular state feedback law $\textbf{K}$ defining the control action as $\textbf{u} = \textbf{Kx}$, the following are true
\begin{enumerate}
    \item The affine subspace defined by
    \begin{align}\label{eqn:sls_subspace}
    \begin{bmatrix}
    I-Z\A&-Z\B
    \end{bmatrix}\begin{bmatrix}
    \Phi_x\\\Phi_u
    \end{bmatrix}= I\;,
    \end{align}
    parametrizes all possible system responses as 
\begin{align}\label{eqn:sls_response}
    \begin{bmatrix}
    \textbf{x}\\\textbf{u}
    \end{bmatrix}=\begin{bmatrix}
    \Phi_x\\\Phi_u
    \end{bmatrix}\begin{bmatrix}
    x_0\\\textbf{w}
    \end{bmatrix}
\end{align}
\item For any block-lower-triangular matrices ${\Phi_x, \Phi_u}$ satisfying~\eqref{eqn:sls_subspace}, the controller $K = \Phi_u\Phi_x^{-1}$ achieves the desired response.
\end{enumerate}
\end{theorem}

\section{Equivalence between Fundamental Lemma and SLS in Uncertain LTI systems}\label{sect:eqv}
For the sake of clarity, the equivalence between the fundamental lemma and the SLS is established by the system~\eqref{eqn:lin_dyn_stc}, its general equivalence for the following uncertain LTI system will be discussed in Remark~\ref{rmk:gen_eqv}.
\begin{align}\label{eqn:lin_dyn_stc_part}
\begin{split}
    x_{i+1} &= Ax_i+Bu_i+Ew_i\\
    y_i &= Cx_i+Du_i+Fw_i\;.
\end{split}
\end{align}
The dimensions of the inputs $u$, the states $x$ and the process noise $w$ are denoted by $n_u$, $n_x$ and $n_w$. Because the disturbance can be considered as an uncontrolled input, the fundamental lemma can be generalized to the system~\eqref{eqn:lin_dyn_stc_part}. Given an augmented input/output sequence $\{\Tilde{u}_i,\Tilde{w}_i,\Tilde{y}_i\}_{i=0}^T$, by defining $\Han_L(\Tilde{y},\Tilde{w},\Tilde{u})$ and $\mathfrak{B}_L(A,B,C,D,E,F)$ in a form similar to the fundamental lemma (Lemma~\ref{lem:funda}), the extension of the fundamental lemma is concluded as follows.
\begin{corollary}\label{cor:funda}
Consider a controllable linear system and assume $\Tilde{u}$ and $\Tilde{w}$ are both persistently exciting of order $L+ \mathfrak{O}(\mathfrak{B}(A,B,C,D,E,F))$. Then $\text{colspan}(\Han_L(\Tilde{y},\Tilde{w},\Tilde{u}))=\mathfrak{B}_L(A,B,C,D,E,F)$. 
\end{corollary}

Consider the fully observed system~\eqref{eqn:lin_dyn_stc}, Corollary~\ref{cor:funda} implies that for each trajectory $\textbf{x}$ driven by inputs $\textbf{u}$ and disturbance $\textbf{w}$, there exists a $g\in \mathbb{R}^{n_c}$ such that 
\begin{align}\label{eqn:funda_subspace}
    \begin{bmatrix}
    \Han_L(\Tilde{u})\\\Han_L(\Tilde{w})\\\Han_{L+1,1}(\Tilde{x})\\\Han_{L+1,2:L+1}(\Tilde{x})
    \end{bmatrix}g = \begin{bmatrix}
    \textbf{u}\\\textbf{w}\\\textbf{x}
    \end{bmatrix}\;,
\end{align}
where $\Han_{L+1}(\Tilde{x})$ is split into two parts as 
\begin{align*}
    \Han_{L+1}(\Tilde{x}):= \begin{bmatrix}
    \Han_{L+1,1}(\Tilde{x})\\\Han_{L,2:L}(\Tilde{x})
    \end{bmatrix}\;.
\end{align*}
In particular, $\Han_{L+1,1}(\Tilde{x}):= [\Tilde{x}_0,\Tilde{x}_1\dots,\Tilde{x}_{n_c-1}]$ are the first row block of $\Han_{L+1}(\Tilde{x})$ corresponding to the initial components. Accordingly, $\Han_{L+1,2:L+1}(\Tilde{x})$ is the second to the $(L+1)$-th row block of $\Han_{L+1}(\Tilde{x})$. It is noteworthy to point out that the Hankel matrix of $\Tilde{x}$ is of depth $L+1$, because system~\eqref{eqn:lin_dyn_stc} has a one-step delay from $u$ to $x$. We conclude the following Lemma
\begin{lemma}\label{lem:eqv}
The subspace defined in~\eqref{eqn:sls_subspace} and~\eqref{eqn:sls_response} is the same as the subspace defined by~\eqref{eqn:funda_subspace}.
\end{lemma}
\begin{proof}
Equation~\eqref{eqn:sls_response} implies that 
\begin{align*}
    \begin{bmatrix}
\textbf{x}\\\textbf{u}
\end{bmatrix}\in \text{colspan}\left(\begin{bmatrix}
\Phi_x\\\Phi_u
\end{bmatrix}\right)\;.
\end{align*}
By Corollary~\ref{cor:funda}, there exists a linear map $G$ such that 
\begin{align*}
    \begin{bmatrix}
\Phi_x\\\Phi_u
\end{bmatrix} = \begin{bmatrix}
\Han_{L+1}(\Tilde{x})\\\Han_L(\Tilde{u})
\end{bmatrix}G\;,
\end{align*}
therefore equation~\eqref{eqn:sls_subspace} is rewritten as
\begin{align}\label{eqn:proof_lem_eqv}
\begin{bmatrix}
I-Z\A&-Z\B 
\end{bmatrix}\begin{bmatrix}
\Han_{L+1}(\Tilde{x})\nonumber\\
\Han_L(\Tilde{u})
\end{bmatrix}G &= I\\\stackrel{(a)}{\Longrightarrow}
\begin{bmatrix}
\Han_{L+1,1}(\Tilde{x})\\\Han_L(\Tilde{w}) 
\end{bmatrix}G & = I\;,
\end{align}
where $(a)$ comes from the substitution of dynamics~\eqref{eqn:lin_dyn_stc}. By denoting $\hat{g}=G[x_0^\top,\textbf{w}]^\top$, we have
\begin{align*}
        \begin{bmatrix}
   x_0\\\textbf{w} 
    \end{bmatrix} =I \begin{bmatrix}
    x_0\\\textbf{w} 
    \end{bmatrix} \stackrel{(b)}{=}  \begin{bmatrix}
\Han_{L+1,1}(\Tilde{x})\\\Han_L(\Tilde{w}) 
\end{bmatrix}G\begin{bmatrix}
x_0\\\textbf{w} 
\end{bmatrix} = \begin{bmatrix}
\Han_{L+1,1}(\Tilde{x})\\\Han_L(\Tilde{w}) 
\end{bmatrix}\hat{g}\;,
\end{align*}
where (b) results from~\eqref{eqn:proof_lem_eqv}. Meanwhile, we have
\begin{align*}
        \begin{bmatrix}
    \textbf{x}\\\textbf{u}
    \end{bmatrix}=\begin{bmatrix}
    \Phi_x\\\Phi_u
    \end{bmatrix}\begin{bmatrix}
    x_0\\\textbf{w}
    \end{bmatrix}=\begin{bmatrix}
\Han_{L+1}(\Tilde{x})\\\Han_L(\Tilde{u})
\end{bmatrix}G\begin{bmatrix}
    x_0\\\textbf{w} 
    \end{bmatrix}= \begin{bmatrix}
\Han_{L+1}(\Tilde{x})\\\Han_L(\Tilde{u})
\end{bmatrix}\hat{g}\;,
\end{align*}
which shows that every element in the subspace~\eqref{eqn:sls_subspace} and~\eqref{eqn:sls_response} corresponds to an element in the subspace~\eqref{eqn:funda_subspace} by linear transformation. Finally, since the feedback law can be arbitrary, the subspace~\eqref{eqn:sls_subspace} and~\eqref{eqn:sls_response} is of dimension $n_x+L(n_u+n_w)$, which is equal to the dimension of subspace~\eqref{eqn:funda_subspace}. Hence, these two subspaces are equivalent.
\end{proof}
\begin{remark}\label{rmk:ctrl_eqv}
Lemma~\ref{lem:eqv} holds intuitively as they both represent the same system. The proof shows the exact link between the fundamental lemma and the SLS. Meanwhile, this Lemma implies that a model-based controller is essentially equivalent to a data-driven control.
\end{remark}
\begin{remark}\label{rmk:gen_eqv}
To generalize the equivalence to system~\eqref{eqn:lin_dyn_stc_part}, we only need to further assume the observability of the system~\eqref{eqn:lin_dyn_stc_part}. The observability essentially implies that the sequence of $\{\textbf{u},\textbf{w},\textbf{y}\}$ uniquely determines the sequence of $\textbf{x}$, a further application of Lemma~\ref{lem:eqv} shows the general equivalence.
\end{remark}

\section{Robust closed-loop DeePC}\label{sect:rb_deepc}
Based on our discussion in Remark~\ref{rmk:ctrl_eqv}, there must exist a data-driven robust controller for system~\eqref{eqn:lin_dyn_stc_part}, which is constructed by augmented input-output data $\{\Tilde{u}_i,\Tilde{w}_i,\Tilde{y}_i\}_{i=0}^T$. According to the Corollary~\ref{cor:funda}, we first modify the prediction part in DeePC~\eqref{eqn:deepc} to  
\begin{align}\label{eqn:rb_pred}
\begin{split}
    \forall\; \textbf{w}_{pred}\in\mathcal{W}:= \{w|F_w w\leq f_w\}\\
    \begin{bmatrix}
    \Han_{L,init}(\Tilde{u})\\\Han_{L,init}(\Tilde{w})\\\Han_{L,init}(\Tilde{y})\\\Han_{L,pred}(\Tilde{u})\\\Han_{L,pred}(\Tilde{w})\\\Han_{L,pred}(\Tilde{y})
    \end{bmatrix}g=\begin{bmatrix}
        \textbf{u}_{init}\\\textbf{w}_{init}\\\textbf{y}_{init}\\\textbf{u}_{pred}\\\textbf{w}_{pred}\\\textbf{y}_{pred}\\
    \end{bmatrix}\;,
    \end{split}
\end{align}
whose elements are all defined in an approach similar to the standard DeePC~\eqref{eqn:deepc} and the disturbance is assumed to be bounded in a polytope. For the sake of compactness, we denote the prediction part in~\eqref{eqn:rb_pred} as
\begin{align*}
    \begin{bmatrix}
    \mathcal{H}_{init}\\\Han_{L,pred}(\Tilde{u})\\\Han_{L,pred}(\Tilde{w})\\\Han_{L,pred}(\Tilde{y})
    \end{bmatrix}g = \begin{bmatrix}
    h_{init}\\\textbf{u}_{pred}\\\textbf{w}_{pred}\\\textbf{y}_{pred}
    \end{bmatrix}
\end{align*}

In this section, we will first formulate a general, but not necessarily causal, data-driven robust controller. Its causal realization, which we coin \textbf{Robust DeePC}, is further introduced in Theorem~\ref{thm:causal_rb}.

\subsection{General Robust DeePC}\label{sect:gen_rb_deepc}
Unlike a feedback controller defined by a state space model, in a DeePC scheme, inputs $\textbf{u}_{pred}$ and $\textbf{y}_{pred}$ are coupled indirectly through $g$. Hence, we propose to define a feedback control law on $g$ and then show that this is equivalent to state feedback. Similar to most feedback laws used in robust MPC, $g$ is decomposed into a nominal part $\overline{g}$ and a linear feedback part $K_d\textbf{w}_{pred}$ as
\begin{align}\label{eqn:fb_g}
    g = \overline{g}+K_d\textbf{w}_{pred}\;.
\end{align}

Based on this control law, a robust data-driven control is stated as
\begin{lemma}\label{lem：rb_cons}
If $\overline{g}$ and $K_d$ satify following constraints, then the control law~\eqref{eqn:fb_g} guarantees $n_h$-step robust feasibility.
\begingroup\makeatletter\def\f@size{9.5}\check@mathfonts
\begin{align}\label{eqn:rb_cons}
\begin{split}
    &\begin{bmatrix}
    \mathcal{H}_{init}\\\Han_{L,pred}(\Tilde{w})
    \end{bmatrix}\overline{g}=\begin{bmatrix}
    h_{init}\\\textbf{0}
    \end{bmatrix}\;,\\
    &\mathcal{H}_{init}
    K_d=\textbf{O}\;,\;
    \Han_{L,pred}(\Tilde{w})K_d = I\;,\\
    &\forall\;\textbf{w}_{pred}\in\mathcal{W}\;,\\
    &\begin{bmatrix}
    \Han_{L,pred}(\Tilde{u})\\\Han_{L,pred}(\Tilde{y})
    \end{bmatrix}(\overline{g}+K_d\textbf{w}_{pred})=\begin{bmatrix}
        \overline{\textbf{u}}_{pred}+\textbf{u}_{fb}\\\overline{\textbf{y}}_{pred}+\textbf{y}_{fb}
    \end{bmatrix}\\
    &\begin{bmatrix}
    F_u&\textbf{O}\\
    \textbf{O}&F_y
    \end{bmatrix}\begin{bmatrix}
        \overline{\textbf{u}}_{pred}+\textbf{u}_{fb}\\\overline{\textbf{y}}_{pred}+\textbf{y}_{fb}
    \end{bmatrix} \leq \begin{bmatrix}
    f_u\\f_y
    \end{bmatrix}\;.
\end{split}
\end{align}
\endgroup
\end{lemma}
\textbf{Proof. } In the control law~\eqref{eqn:fb_g}, the nominal $\overline{g}$ generates a disturbance-free $n_h$ step prediction. Hence, based on the prediction equation~\eqref{eqn:rb_pred}, we enforce 
\begingroup\makeatletter\def\f@size{9.5}\check@mathfonts
\begin{align}\label{eqn:pred_pf}
    \begin{split}
        \forall\;\textbf{w}_{pred}&\in\mathcal{W}\;\\
        \begin{bmatrix}
            \mathcal{H}_{init}\\\Han_{L,pred}(\Tilde{u})\\\Han_{L,pred}(\Tilde{w})\\\Han_{L,pred}(\Tilde{y})
        \end{bmatrix}&(\underbrace{\overline{g}}_{(a)}+\underbrace{K_d\textbf{w}_{pred}}_{(b)})=\begin{bmatrix}
        h_{init}\\\overline{\textbf{u}}_{pred}\\\textbf{0}\\\overline{\textbf{u}}_{pred}
        \end{bmatrix}+\begin{bmatrix}
        \textbf{0}\\\textbf{u}_{fb}\\\textbf{w}_{pred}\\\textbf{y}_{fb}
        \end{bmatrix}
    \end{split}\;,
\end{align}
\endgroup
where the matrix products of $(a)$ and $(b)$ correspond to the components on the right-hand side accordingly. As the future disturbance $\textbf{w}_{pred}$ is unknown and arbitrary within the polytope $\mathcal{W}$, the matrix product of term $(b)$ in~\eqref{eqn:pred_pf} implies
\begin{align}\label{eqn:fb_prop}
    \mathcal{H}_{init}K_d=\textbf{O}\;,\;
    \Han_{L,pred}(\Tilde{w})K_d = I.
\end{align}
Due to the perturbation of the unknown future disturbance, the actual input and the actual output under the control law~\eqref{eqn:fb_g} are $\overline{\textbf{u}}_{pred}+\textbf{u}_{fb}$ and $\overline{\textbf{y}}_{pred}+\textbf{y}_{fb}$ respectively, which give the the robust constraints in~\eqref{eqn:rb_cons}. Hence, we conclude the proof.
\hfill$\blacksquare$

Lemma~\ref{lem：rb_cons} allows us to define a robust data-driven control problem 
\begingroup\makeatletter\def\f@size{9}\check@mathfonts
\begin{align}\label{eqn:gen_rb_deepc} 
\begin{split}
\min\limits_{\substack{\overline{g},K_d,\\\textbf{u}_{pred},\textbf{y}_{pred}}}&\max\limits_{\textbf{w}_{pred} \in \mathcal{W}}\;\; J(\textbf{y}_{pred},\textbf{u}_{pred})\\
    \text{s.t}&\begin{bmatrix}
    \mathcal{H}_{init}\\\Han_{L,pred}(\Tilde{w})
    \end{bmatrix}\overline{g}=\begin{bmatrix}
    h_{init}\\\textbf{0}
    \end{bmatrix}\;,\\
    &\mathcal{H}_{init}
    K_d=\textbf{O}\;,\;
    \Han_{L,pred}(\Tilde{w})K_d = I\;,\\
    &\forall\;\textbf{w}_{pred}\in\mathcal{W}\;,\\
    &\begin{bmatrix}
    \Han_{L,pred}(\Tilde{u})\\\Han_{L,pred}(\Tilde{y})
    \end{bmatrix}(\overline{g}+K_d\textbf{w}_{pred})=\begin{bmatrix}
        \overline{\textbf{u}}_{pred}+\textbf{u}_{fb}\\\overline{\textbf{y}}_{pred}+\textbf{y}_{fb}
    \end{bmatrix}\\
    &\begin{bmatrix}
    F_u&\textbf{O}\\
    \textbf{O}&F_y
    \end{bmatrix}\begin{bmatrix}
        \overline{\textbf{u}}_{pred}+\textbf{u}_{fb}\\\overline{\textbf{y}}_{pred}+\textbf{y}_{fb}
    \end{bmatrix} \leq \begin{bmatrix}
    f_u\\f_y
    \end{bmatrix}\;.
\end{split}
\end{align}
This problem is a standard robust optimization problem, which can be reformulated as a convex optimization problem with a dualization technique~\cite{ben2009robust}. To clarify this procedure, we define
\begin{align*}
    \begin{bmatrix}
    \Tilde{f}_u\\\Tilde{f}_y
    \end{bmatrix}:=\begin{bmatrix}
    f_u\\f_y
    \end{bmatrix}- \begin{bmatrix}
    F_u&\textbf{O}\\
    \textbf{O}&F_y
    \end{bmatrix}\begin{bmatrix}
        \Han_{L,pred}(\Tilde{u})\\\Han_{L,pred}(\Tilde{y})
    \end{bmatrix}\overline{g};.
\end{align*}
The resulting convex optimization based on the dualization technique is 
\begin{align}
    \min\limits_{\substack{\overline{g},K_d,\Lambda\\\textbf{u}_{pred},\textbf{y}_{pred}}}& \;\; J(\textbf{y}_{pred},\textbf{u}_{pred})\nonumber\\
    \text{s.t.}\;&\;\begin{bmatrix}
    \mathcal{H}_{init}\\\Han_{L,pred}(\Tilde{w})\\\Han_{L,pred}(\Tilde{u})\\\Han_{L,pred}(\Tilde{w})
    \end{bmatrix}\overline{g}=\begin{bmatrix}
    h_{init}\\\textbf{0}\\\overline{\textbf{u}}_{pred}\\\overline{\textbf{y}}_{pred}
    \end{bmatrix}\;,\nonumber\\
    &\mathcal{H}_{init}
    K_d=\textbf{O}\;,\;
    \Han_{L,pred}(\Tilde{w})K_d = I\;,\nonumber\\
    & \Lambda^\top f_w \leq \begin{bmatrix}
    \Tilde{f}_u\\\Tilde{f}_y
    \end{bmatrix}\;,\nonumber\\
    & F_w^\top \Lambda = K_d^\top\begin{bmatrix}
    F_u&\textbf{O}\\\textbf{O}&F_y
    \end{bmatrix}^\top\;,\nonumber\\
    & \Lambda \geq \textbf{O}\label{eqn:cons_lambda}\;,
\end{align}
where constraint~\eqref{eqn:cons_lambda} is imposed element-wise. The dual variable matrix is $\Lambda\in\mathbb{R}^{n_w\times n_{neq}}$ with $n_{neq}$ the total number of inequality constraints imposed on inputs and outputs, each column of $\Lambda$ corresponds to the dual variable of one inequality constraint of the outputs or the inputs.

\subsection{Causal Robust DeePC}

Before discussing the details of this section, we first recall some notation of the cardinalities used in the previous sections. $n_y$, $n_u$ and $n_w$ are the dimensions of the inputs, the outputs and the process noise, $n_c$ is the number of columns in the Hankel matrices, $n_h$ is the prediction horizon in the optimal control problem~\eqref{eqn:gen_rb_deepc} and $t_{init}$ is the depth used in the initialization Hankel matrix. Based on this, we define $\Han_{L,i}(\cdot)$ to be the $i$-th row block of the Hankel matrix $\Han_{L}$, for example $\Han_{L,i}(\Tilde{x})=[\Tilde{x}_i,\Tilde{x}_{i+1}\dots,\Tilde{x}_{i+n_c-1}]$ and the $i$-th row block of  $\Han_{L,pred,i}(\Tilde{x})$ is $[\Tilde{x}_i+t_{init},\Tilde{x}_{i+1+t_{init}}\dots,\Tilde{x}_{i+n_c-1+t_{init}}]$. For the sake of compactness, we further use the \textsc{Matlab} index notation, such that $\Han_{L,i:j}$ is the $i$-th to $j$-th row block of $\Han_{L}$ and the $i$-th to the $j$-th measurement of sequence $\textbf{x}$ is $\textbf{x}_{i:j}= [x_i^\top,x_{i+1}^\top \dots,x_j^\top]^\top$. Finally, we define $K_{d,:,j}$ as the $j$-th block column of feedback law $K_d$, which corresponds to the feedback generated by $\textbf{w}_{pred,i}$ and that is the $(n_w\times (i-1))+1$-th to $n_w\times i$-th column of $\textbf{w}_{pred}$

Here starts the main result of this section. If the feedback matrix $K_d$ is arbitrary, then the feedback control law is not necessarily causal. In particular, the feedback computed from $\textbf{w}_{pred,i}$ should not be able to change $\textbf{u}_{pred,1:i}$ and $\textbf{y}_{pred,1:i-1}$, because those events happen before $\textbf{w}_{pred,i}$. 

Consider now a causal linear feedback control law on the disturbance
\begin{align*}
    \textbf{u}_{pred} = \overline{\textbf{u}}_{pred}+K_w \textbf{w}_{pred}\;,
\end{align*}
where the feedback law $K_w$ has a causal structure as
\begin{align*}
    K_w:=\begin{bmatrix}
    \textbf{O}&&&\\
    K_{w,1,0}&\textbf{O}&&\\
    \vdots&\ddots&\ddots&\\
    K_{w,n_h,0}&\dots&K_{w,n_h,n_h-1}&\textbf{O}
    \end{bmatrix}\;.
\end{align*}
The sub-matrices $K_{w,i,j}$ are of size $\mathbb{R}^{n_u\times n_w}$. We define a standard robust MPC controller based on this causal control law as
\begin{align}\label{eqn:rb_mpc}
\begin{split}
    \min\limits_{\substack{\textbf{u}_{pred},K_w\\\textbf{y}_{pred}}} &\max_{\textbf{w}_{pred}}J(\textbf{y}_{pred},\textbf{u}_{pred})\\
    \text{s.t.}\;& \textbf{x}_{pred,0} = \textbf{x}_{t_{init}}\\
    &\forall\; i = 0,1\dots,n_h\\
    &\textbf{x}_{pred,i+1} = A\textbf{x}_{pred,i}+B\textbf{u}_{pred,i}+E\textbf{w}_{pred,i}\\
    & \textbf{y}_{pred,i} = C\textbf{x}_{pred,i}+D\textbf{u}_{pred,i}+F\textbf{w}_{pred,i}\;\\
    & \forall\; \textbf{w}_{pred}\in\mathcal{W}\\
    & \textbf{u}_{pred}=\overline{\textbf{u}}_{pred}+K_w\textbf{w}_{pred}\\
    & F_u\textbf{u}_{pred}\leq f_u\; F_y\textbf{y}_{pred}\leq f_y\;,
\end{split}
\end{align}
where $\textbf{u}_{pred}:=[u_{pred,0}^\top,\dots,u_{pred,n_h}^\top]$,  $\textbf{y}_{pred}$ and $\textbf{w}_{pred}$ are defined accordingly. The goal of this section is to design a data-driven robust controller, whose resulting control law is identical to the model-based controller~\eqref{eqn:rb_mpc}. 

To construct the causal data-driven control law, we define
\begin{align}
    H=\begin{bmatrix}
    \Han_{L,init}(\Tilde{u})\\\Han_{L,init}(\Tilde{w})\\\Han_{L,init}(\Tilde{y})\\\Han_{L,pred,1}(\Tilde{u})\\\Han_{L,pred,1}(\Tilde{y})\\\vdots\\\Han_{L,pred,n_h}(\Tilde{u})\\\Han_{L,pred,n_h}(\Tilde{y})
    \end{bmatrix}\;,
\end{align}
with $H\in\mathbb{R}^{n_r\times n_c}$ and $n_r$ the number of rows. The QR decomposition~\cite{stewart1998matrix} of its transpose is
\begin{align*}
    H^\top = \begin{bmatrix}
    Q_a&Q_b
    \end{bmatrix}\begin{bmatrix}
    R\\\textbf{O}
    \end{bmatrix}\;,
\end{align*}

\begin{theorem}\label{thm:causal_rb}
If the system~\eqref{eqn:lin_dyn_stc_part} is observable, then the model-based robust control law in~\eqref{eqn:rb_mpc} is identical to the following data-driven control law,
\begingroup\makeatletter\def\f@size{9.5}\check@mathfonts
\begin{align}\label{eqn:rb_deepc_full}
\begin{split}
    \min\limits_{\substack{\overline{g},K_p,\\\textbf{u}_{pred},\textbf{y}_{pred}}}&\max\limits_{\textbf{w}_{pred} \in \mathcal{W}}\;\; J(\textbf{y}_{pred},\textbf{u}_{pred})\\
    \text{s.t.}\;&\;\begin{bmatrix}
    \mathcal{H}_{init}\\\Han_{L,pred}(\Tilde{w})
    \end{bmatrix}\overline{g}=\begin{bmatrix}
    h_{init}\\\textbf{0}
    \end{bmatrix}\\
    &K_d = \begin{bmatrix}
    Q_{a,:,n_{init}+1:n_r}&Q_b
    \end{bmatrix}K_p\;,\\
    &\mathcal{H}_{init}
    K_d=\textbf{O}\;,\;
    \Han_{L,pred}(\Tilde{w})K_d = I\\
    &\forall\;\textbf{w}_{pred}\in\mathcal{W}\;\\
    &\begin{bmatrix}
    \Han_{L,pred}(\Tilde{u})\\\Han_{L,pred}(\Tilde{y})
    \end{bmatrix}(\overline{g}+K_d\textbf{w}_{fb})=\begin{bmatrix}
        \overline{\textbf{u}}_{pred}+\textbf{u}_{pred}\\\overline{\textbf{y}}_{pred}+\textbf{y}_{fb}
    \end{bmatrix} \\
    &\begin{bmatrix}
    F_u&\textbf{O}\\
    \textbf{O}&F_y
    \end{bmatrix}\begin{bmatrix}
        \overline{\textbf{u}}_{pred}+\textbf{u}_{fb}\\\overline{\textbf{y}}_{pred}+\textbf{y}_{fb}
    \end{bmatrix} \leq \begin{bmatrix}
    f_u\\f_y
    \end{bmatrix}
\end{split}
\end{align}
\endgroup
where $n_{init}:=t_{init}\times(n_u+n_y)+n_u$ and $K_p$ has a lower block triangular structure as
\begin{align*}
    K_p = \begin{bmatrix}
    K_{p,1,1} &\textbf{O}&\textbf{O}&\dots&\textbf{O}\\
    K_{p,2,1} & K_{p,2,2}&\textbf{O}&\dots&\textbf{O}\\
    \vdots&\vdots&\ddots&\ddots&\vdots\\
    K_{p,n_h,1}&K_{p,n_h,2}&K_{p,n_h,3}&\dots&K_{p,n_h,n_h}
    \end{bmatrix}\;.
\end{align*}
$K_{p,i,j}$ are dense matrix blocks, whose sizes are $(n_u+n_y)\times n_w$ for $\forall\; i<n_h$ and are $[n_y+(n_c-n_r)]\times n_w$ when $i=n_h$.
\end{theorem}
\textbf{Proof.} First, it is observed that the data-driven formulation is based on the robust controller~\eqref{eqn:gen_rb_deepc}, where $\overline{\textbf{y}}_{pred}$ and $\overline{\textbf{u}}_{pred}$ are both well-defined. By Lemma~\ref{lem:eqv} and our discussion in Remark~\ref{rmk:ctrl_eqv}, the control law will be equivalent once causality is enforced.

We recall a useful property of QR decomposition~\cite{stewart1998matrix}: the range of the first $n$ rows of $H$ is spanned by the first $n$ columns of $Q_a$. Because $[Q_a,Q_b]$ is an unitary matrix, the remaining $n_r-n$ columns in $Q_a$ and the matrix $Q_b$ forms the null space of the first $n$ rows in matrix $H$. Considering the constraint $\mathcal{H}_{init}K_d = \textbf{O}$ in problem~\eqref{eqn:gen_rb_deepc}, each column of matrix $K_d$ must lie within the null space of $\mathcal{H}_{init}$. Meanwhile, the feedback ingredients from $\textbf{w}_{pred}$ cannot change the value of the first input $\textbf{u}_{pred,0}$. In conclusion, we enforce
\begin{align*}
    \text{colspan}(K_d) &\subset \text{colspan}([Q_{a,:,n_{init}+1:n_r},Q_b])\;,
\end{align*} 
with $n_{init}:=t_{init}\times(n_u+n_y)+n_u$. The $j$-th column in matrix $K_d$, $K_{d,:,j}$, defines the feedback ingredient with respect to $\textbf{w}_{pred,j}$. By causality, the feedback from $\textbf{w}_{pred,j}$ should not be able to change the inputs and outputs that occur before $\textbf{w}_{pred,j}$. In particular, $K_{d,:,j}$ should further lie in the null space of the matrices $\Han_{L,pred,1:i}(\Tilde{u}),\Han_{L,pred,1:i-1}(\Tilde{y})$, we therefore enforce
\begin{align*}
    K_{d,:,j}\subset \text{colspan}([Q_{a,:,n_{init}+(j-1)(n_u+n_y)+1:n_r},Q_b])\;.
\end{align*}
All the constraints on the null spaces can be reformulated as
\begin{align*}
    K_d = \begin{bmatrix}
    Q_{a,:,n_{init}+1:n_r}&Q_b
    \end{bmatrix}K_p\;,
\end{align*}
which concludes the proof.
\hfill $\blacksquare$

We call the controller proposed in Theorem~\ref{thm:causal_rb} a \textbf{robust data-enabled predictive controller (robust DeePC)}. The robust DeePC problem can also be reformulated into a convex optimization problem with the dualization technique discussed in the Section~\ref{sect:gen_rb_deepc}.

\subsection{Discussion}
In comparison with a model-based robust controller~\eqref{eqn:rb_mpc}, the proposed controller distinguishes itself by a data-driven convex formulation. Meanwhile, the proposed scheme has the same scale of computational cost. In particular, the size of the optimization problem~\eqref{eqn:rb_deepc_full} only differs in the formulation of the feedback, where the number of decision variables in the feedback control law is $O(n_h\times(n_u+n_y))$ due to the causal reformulation. Hence, the computational cost of the robust DeePC control scheme is similar to the robust MPC.

In terms of the online data-driven control, the computational cost of the Hankel matrix update is low. In particular, the computational cost of the QR decompostion update by adding or removing a column scales linearly with respect to the size of the Hankel matrix~\cite[Section 6.5]{golub2013matrix}.

\section{Numerical Example}\label{sect:demos}
In this section, numrical experiments are carried out to validate the proposed robust DeePC. First, a second order system is used to show the equivalence between the proposed robust DeePC and a the MPC~\eqref{eqn:rb_mpc} with full state measurement. After that, we test the proposed scheme in a building control problem to adapt power consumption with respect to the occupation patterns. The code is implemented with \textsc{Yalmip}~\cite{Lofberg2004} interfacing the \textsc{Gurobi} solver~\cite{gurobi}
\subsection{Second Order System}
The proposed scheme is compared against a robust MPC controller~\eqref{eqn:rb_mpc} , and is tested on a second order system:
\begingroup\makeatletter\def\f@size{9}\check@mathfonts
\begin{align*}
    x_{i+1} &= \begin{bmatrix}
    0.9535  &  0.0761\\-0.8454  &  0.5478
    \end{bmatrix}x_i+\begin{bmatrix}
    0.0465\\ 0.8454
    \end{bmatrix}u_i+\begin{bmatrix}
    0.0465\\0.8454
    \end{bmatrix}w_i\\
    y_i&=\begin{bmatrix}
    1&0
    \end{bmatrix}x_i\;,
\end{align*}
\endgroup
the process noise $w$ is bounded within $[-0.1,0.1]$, the inputs and outputs are constrained by $u\in[-5,5],y\in[-0.5,0.5]$. A quadratic stage cost is used 
\begin{align*}
    J(\textbf{y}_{pred},\textbf{u}_{pred})&=\sum\limits_{i=1}^{n_h}(\textbf{y}_{pred,i+1}
    -r)^\top Q(\textbf{y}_{pred,i+1}-r)\\&+\textbf{u}_{pred,i}^\top R \textbf{u}_{pred,i}\;,
\end{align*} 
where $Q = 10$, $R = 0.1$ and $r$ is the reference. The Hankel matrices in the robust DeePC are built with a sequence of length $100$. The tracking performance of the proposed controller is shown in Figure~\ref{fig:compare}, where the robust MPC has full state measurement. As we claimed in Section~\ref{sect:rb_deepc}, the response of the proposed controller is the same as the robust MPC~\eqref{eqn:rb_mpc} with full state measurement. The controller can safely protect the system away from the constraint by considering the perturbation caused by the future disturbance.
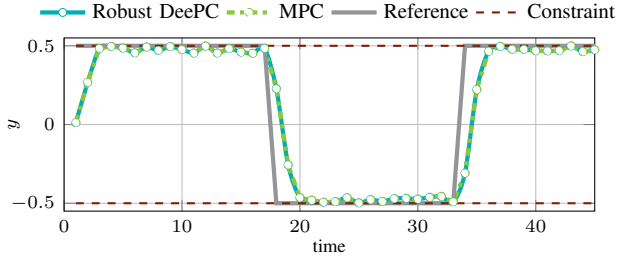
\begin{figure}[h]
    \centering
    \begin{tikzpicture}
    \begin{axis}[xmin=0, xmax=45,
    ymin=-0.55, ymax= 0.55,
    enlargelimits=false,
    clip=true,
    grid=major,
    mark size=0.5pt,
    width=1\linewidth,
    height=0.45\linewidth,ylabel = $y$,xlabel= time,
    legend style={
    	font=\footnotesize,
    	draw=none,
		at={(0.5,1.03)},
        anchor=south
    },
    ylabel style={at={(axis description cs:0.08,0.5)}},
    xlabel style={at={(axis description cs:0.5,0.12)}},
    legend columns=4,
    label style={font=\scriptsize},
    ticklabel style = {font=\scriptsize}]
    
    \pgfplotstableread{data/2dsys.dat}{\dat}
    
    \addplot+ [ultra thick, smooth, mark=*, mark options={fill=white, scale=3,line width = 0.2pt}, TealBlue] table [x={t}, y={deepc}] {\dat};
    \addlegendentry{Robust DeePC}
  
  \addplot+ [ultra thick,dashdotted, mark=*, mark options={fill=white, scale=3,line width = 0.1pt},LimeGreen] table [x={t}, y={mpc}] {\dat};
    \addlegendentry{MPC}
    
    \addplot+ [ultra thick,Gray,mark = none] table [x={t}, y={ref}] {\dat};
    \addlegendentry{Reference}
    
    \addplot+ [thick,Brown,dashed,mark = none] table [x={t}, y={max}] {\dat};
    \addplot+ [thick,Brown, dashed, mark = none] table [x={t}, y={min}] {\dat};
    \addlegendentry{Constraint}

    \end{axis}
    \end{tikzpicture}
    \caption{Comparison of the Robust DeePC and the MPC controller. Note that the two responses are the same.}
    \label{fig:compare}
\end{figure}

\subsection{Building Control}
We consider a single zone building model, which is disturbed by internal heat gain, solar radiation and external temperature. The model used to generate the data is
\begin{align*}
    x_{i+1} &= \begin{bmatrix}
    0.8511 & 0.0541 & 0.0707\\
    0.1293 & 0.8635 & 0.0055\\
    0.0989 & 0.0032 & 0.7541
    \end{bmatrix}x_i+\begin{bmatrix}
    0.0035\\0.0003\\ 0.0002
    \end{bmatrix}u_i\\
    &\quad +10^{-3}\begin{bmatrix}
    22.2170 & 1.7912 & 42.2123\\
    1.5376 & 0.6944 & 2.29214\\
    103.1813 & 0.1032 & 196.0444
    \end{bmatrix}w_i\;,\\
    y_i& = \begin{bmatrix}
    1&0&0
    \end{bmatrix}x_i\;,
\end{align*}
where $x$ models the indoor temperature, wall temperature and the corridor temperature respectively. In a building control problem, the controller is designed to maintain occupant comfort while minimizing energy consumption. During the heating season, the indoor temperature is kept above $23^\circ C$ to maintain occupant comfort during the day. When the room is not used at night, the room temperature is only required stay above $17^\circ C$. Beyond the control requirements, the disturbances also show a time dependent pattern. Without loss of generality, we assume that during the day, the solar radiation and the internal heat gain are bounded within $[4,6]$ with an external temperature fluctuating between $[6^\circ C,8^\circ C]$. During the night, the solar radiation is $0$ with much lower internal heat gain ranging between $[0,2]$. Meanwhile, the external temperature is also lower at around $[2^\circ C, 4^\circ C]$. As the controller is designed to minimize power consumption, the loss function is
\begin{align*}
    J(\textbf{y}_{pred},\textbf{u}_{pred})=\lVert\textbf{u}_{pred,i}\rVert_1\;.
\end{align*}

By building all the relevant Hankel matrices with a $100$-step measurement sequence, the performance of the proposed controller is shown in Figure~\ref{fig:building}, where the operation starts from $6$ A.M, and it is already overheated at that point before the controller effectively lowers the indoor temperature. It is also observed that the controller pre-heats the room to slightly above $23^\circ C$ before the $6$ A.M before the second morning. This whole cycle shows the effectiveness of the controller.
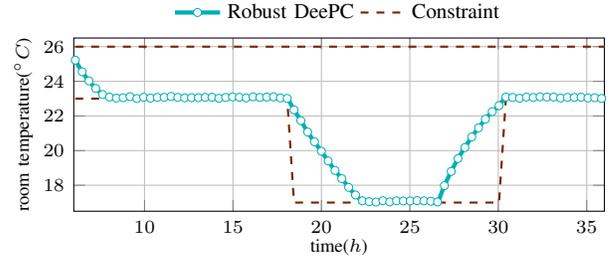
\begin{figure}[h]
    \centering
    \begin{tikzpicture}
    \begin{axis}[xmin=6, xmax=36,
    ymin=16.5, ymax= 26.5,
    enlargelimits=false,
    clip=true,
    grid=major,
    mark size=0.5pt,
    width=1\linewidth,
    height=0.45\linewidth,ylabel = room temperature($^\circ C$),xlabel= time($h$),
    legend style={
    	font=\footnotesize,
    	draw=none,
		at={(0.5,1.03)},
        anchor=south
    },
    ylabel style={at={(axis description cs:0.08,0.5)}},
    xlabel style={at={(axis description cs:0.5,0.12)}},
    legend columns=2,
    label style={font=\scriptsize},
    ticklabel style = {font=\scriptsize}]
    
    \pgfplotstableread{data/build.dat}{\dat}
    
    \addplot+ [ultra thick, smooth,TealBlue, mark=*, mark options={ scale=3, fill=white,line width=0.1pt}] table [x={t}, y={deepc}] {\dat};
    \addlegendentry{Robust DeePC}

    \addplot+ [thick,Brown,dashed,mark = none] table [x={t}, y={max}] {\dat};
    \addplot+ [thick,Brown, dashed, mark = none] table [x={t}, y={min}] {\dat};
    \addlegendentry{Constraint}

    \end{axis}
    \end{tikzpicture}
    \caption{Temperature control with rboust DeePC}
    \label{fig:building}
\end{figure}

\section{Conclusion}
In this work, we show the equivalence between the SLS and the Willems' fundamental lemma for uncertain LTI systems. A convex data-driven controller is further proposed, which has a control law identical to the robust MPC. A toy example is used to show the equivalence between the robust DeePC and the robust MPC. The robust DeePC is further validated through a building control problem, which maintains occupants' comfort with minimal power consumption.

{\tiny
\bibliographystyle{abbrv}
\bibliography{ref.bib}}

\end{document}